# Room temperature interlayer exciton valley polarization and valley Hall effect

Zumeng Huang[1], Yuanda Liu[1], Kévin Dini[1], Zhuojun Liu[2], Hanlin Fang[2], Jin Liu[2], Timothy Liew[1] and Weibo Gao[1, 3]

*[1]Division of Physics and Applied Physics, School of Physical and Mathematical Sciences, Nanyang Technological University, Singapore*

*[2]State Key Laboratory of Optoelectronic Materials and Technologies, School of Physics, Sun Yat-sen University, Guangzhou 510275, China*

*[3]The Photonics Institute and Centre for Disruptive Photonic Technologies, Nanyang Technological University, Singapore*

**Abstract:**

**For monolayer transition metal chalcogenides (TMDs), electrons and excitons in different valleys can be driven to opposite directions by the Berry curvature[1, 2], serving as a valley-dependent effective magnetic field[3]. In addition to monolayer TMDs, Van der Waals heterostructures provide an attractive platform for emerging valley physics and devices with superior physics properties[4-6]. Interlayer excitons in TMD heterostructures have a long valley lifetime as compared to monolayer intralayer excitons[6]. Here we report an interlayer exciton valley polarization and valley Hall effect in MoS2/WSe2 in room temperature. The separation for excitons with different valley index is observed with polarization-dependent photoluminescence mapping. The exciton separation is perpendicular to their transport directions. The valley Hall effect for indirect excitons is sustained even at room temperature, in contrast with the cryo-temperatures in previous experiments in monolayer TMDs[1, 2]. Room temperature demonstration of the indirect exciton valley polarization and valley Hall effect might open new perspectives for the development of opto-valleytronic devices based on TMD heterostructures.**

Valleys, referring to energy extreme points in momentum space, act as an additional degree of freedom besides spins and charges[7-11]. In recent years, they have attracted great attention due to the ability to carry information and develop valleytronic devices[12]. In monolayer TMD material, two valleys are well separated in the first Brillouin zone with degenerate energy but opposite Berry curvatures. For different valley index, these Berry curvatures serve as opposite orbital magnetic moments for electrons and excitons[13]. Therefore, electrons and excitons from different valleys will move to opposite directions perpendicular to their transport current. This so called valley Hall effect (VHE) is a phenomenon similar to the spin Hall effect[14]. The VHE was first experimentally demonstrated for electrons in MoS2 transistors. For quasi-particles like excitons, VHE was theoretically proposed[15-19] and experimentally demonstrated in monolayer MoS2 with a larger Hall angle[2]. These attributes could open the way towards next generation valleytronics and opto-electronic devices based on two dimensional (2D) materials.

To utilize the valley properties as an information carrier we want a robust valley polarization and a long valley lifetime. TMD valleytronics are believed to have a such properties as the spin-valley locking assures that the intervalley scattering between free electrons and holes will also involves a spin-flip process, which makes the scattering unlikely to happen. However, in PL measurement, the exciton shows a short valley lifetime which is in picosecond range. This is because that the excitons in both valleys will have electron-hole exchange interactions between them and open a channel for valley depolarization.

One of the methods to overcome this fast exciton valley lifetime is to fabricate TMD Van der Waals heterostructures[4, 5]. As the surface of 2D materials is free of dangling bonds, they can be stacked to from such layered system, which provide even richer platforms and bring the possibility for material engineering to exploring their exotic properties. For TMD heterostructures assembled with type-II alignment, electrons and holes favor to travel and stay in different layers with lower energies. This charge transfer process is very fast, typically on the order of 50 fs[20]. With Coulomb interaction, electrons and holes in different layers can form interlayer excitons[21]. These excitons, with a reduced electron and hole wavepacket overlap, are less likely for recombination and depolarization through the exchange interaction, and have a longer exciton lifetime and valley lifetime comparing with intralayer excitons in monolayers[6, 22]. Therefore, they are ideal for exciton to carry valley information and travel a longer distance[23]. In addition, interlayer excitons have an out-of-plane permanent electric dipole moment, which is similar to indirect excitons in traditional GaAs/AlGaAs quantum wells[24] and makes them proper candidates for electrical control towards exciton switches[25-27]. All these properties make them interesting candidates for studying valley physics and fabricating valleytronic devices.

Here we report a room temperature VHE for interlayer excitons in MoS2/Wse2 heterostructures. Due to the type-II band alignment for MoS2 and WSe2, electrons and holes will transfer to MoS2 and WSe2 in ultra-short time, respectively[20, 28, 29]. Our sample has a stacking angle near zero degrees and interlayer excitons bound by electrons and holes in different layers can radiatively recombine and emit photons (See Fig. 1a). With the broken inversion symmetry, the conduction band minima and valence band maxima at $K$ and $K'$ points have a degenerate energy but are inequivalent in momentum space, referred to as their valley degree of freedom. As shown in Fig. 1b, excitons in different valleys carry opposite Berry curvatures and have valley-contrasting transverse motion. In addition, circularly polarized light is coupled with valleys[7, 13, 30] (Fig. 1a), allowing the detection of excitons in different valleys with their exclusive optical helicity ($\sigma^+$ or $\sigma^-$).

Fig. 1c shows the optical microscope image of the real device. WSe2 and MoS2 monolayers were mechanically exfoliated from bulk single crystals onto polydimethylsiloxane (PDMS) stamps. Then they were stacked onto a silicon-on-insulator wafer with 220 nm thick silicon film on a 2 μm sacrificial silicon dioxide layer (See the Methods section for sample preparation details). In the middle of the substrate, an air-bridged silicon photonic crystal slab was fabricated by removing the sacrificial layer with hydrofluoric acid etching. We characterized the MoS2-WSe2 monolayers and heterostructures using photoluminescence (PL) measurements. Their spectrum is shown in Figure 1d. Monolayer MoS2 shows a PL emission with peak wavelength at ~675nm, while WSe2 shows a PL peak centered at ~750nm. In the heterostructure area, a new peak ~1130nm can be seen, which comes from interlayer exciton emission. Several samples have been fabricated to confirm this result (See supplementary section A). We note that our results are consistent with the previous experimental measurement of the bandgap for $K-K$ transitions by scanning microscope spectroscopy[29], but different from $K-\Gamma$ transitions in previous experimental observations[31, 32].

We also investigated the lifetime and valley polarization of such kind of heterostructure (Fig. 1e). It clearly shows an exciton lifetime ~ ns and intensity difference between the co- and cross-polarized emission, for both cryogenic and room temperature. As we extracted the co- and cross-polarized PL intensity data, valley polarization can be calculated as $\frac{C(co)-C(cross)}{C(co)+C(cross)}$, where $C(co)$ and $C(cross)$ are co-polarized and cross-polarized PL counts. Initially at low temperature the valley polarization is about 35%. When temperature goes higher, valley polarization becomes smaller and ends up to around 5%. The noticeable room temperature valley polarization confirms the valley index as a good degree of freedom even up to room temperature.

The photonic crystal suspended slab helps for observing the interlayer exciton VHE. The reason to have these suspended slabs is that they can enhance the PL on them (See supplementary section B) and also create strain at the edge of the photonic crystal area to form local energy traps for the excitons. The strain at the edge will change the bandgap as confirmed by spectrum measurement and induce the exciton funneling effect to attract excitons to the strained area[33, 34]. This introduces asymmetry in the exciton transport direction, which is necessary to observe exciton VHE[16]. In contrast, in previous experiment on intralayer excitons in MoS2, exciton transport is driven by laser illumination and asymmetry is introduced by exciting on the edge of the sample[2].

The experimental setup is shown in Figure 2a and a detailed description can be found in the Methods section. In order to measure the interlayer exciton transport, we perform the PL emission mapping by fixing the excitation laser spot on the sample, and scanning the detection optical path to map the emission pattern at the same time. Note that this is different from the normal PL mapping where the excitation spot and emission spot always overlap with each other in a confocal microscope, as used in our supplementary S1 and S3. To characterize our setup, we first scan the emission pattern for the silicon PL from a heavily doped $SiO_2$-Si wafer which should be centrosymmetric. As expected, the intensity distribution shows a Gaussian shape around the center of the excitation beam (Fig. 2b). Next we proceed to measure the emission pattern for the interlayer excitons. Here we use CW laser excitation with 726nm wavelength and only detect emission from interlayer exciton recombination with long pass filters. In stark contrast with emission from silicon, the emission pattern for interlayer excitons shows a clear distortion from the Gaussian shape (Fig. 2c). Except the strong emission from the middle point (excitation spot), clear sideband emission can be seen at the edges. By comparing the emission pattern with sample images, we see that the suspended slab edge serves as local traps for excitons. The exciton traps here direct the transport and make the VHE observable as demonstrated below. Similar pattern has been observed with different sample substrate without photonic crystal mirror but only strain at the edge as well (shown in supplementary section H).

With a clearly detectable exciton transport, we can then proceed for the polarization dependent PL experiments. Here we use a vertical linear polarization excitation to make sure both valleys are equally populated and first conduct the experiment at 4K. We first scan interlayer exciton PL emission pattern for right ($\sigma^+$) and left ($\sigma^-$) circularly polarized light. With these data, we can then calculate the degree of polarization (DOP) as $\dfrac{C(\sigma^+) - C(\sigma^-)}{C(\sigma^+) + C(\sigma^-)}$ , where $C(\sigma^+)$ and $C(\sigma^-)$ are $\sigma^+$ and $\sigma^-$ PL counts. As can be seen from Fig. 3a, following the transport direction (as indicated by the arrow), we can see a separation between the negative DOP (blue color) and the positive DOP (red color). In stark contrast, the emission of silicon shows a DOP very close to zero (See Fig. S4).

To see it more clearly, Fig. 3b and Fig. 3c show the linecut of PL intensity and DOP following the green dashed line in Fig. 3a. As shown, the intensity peak for $\sigma^+$ and $\sigma^-$ emission is separated along the transverse direction (y). Correspondingly, the DOP sign reverses in different side of the transverse direction. The separation in the emission pattern shows that the interlayer exciton turns right in one valley, resulting in $\sigma^+$ emission, and turns left in the other valley, resulting in $\sigma^-$ emission. The PL intensity and DOP in another direction (yellow linecut) are shown in Fig. 3d and Fig. 3e, which also clearly show the exciton VHE. From Fig.3b and Fig.3d we get the distance between the separated intensity peaks ($\sigma^+$ and $\sigma^-$) to be $0.41 \pm 0.02\,\mu m$ and $0.33 \pm 0.07\,\mu m$ for green and yellow linecut.

Different from the previous VHE in monolayer MoS2, the interlayer exciton VHE can be sustained even at room temperature. In Fig. 4, we have shown the experiment results measured at room temperature with the similar method as described above. As can be seen from Fig. 4a-c, there is a clear separation between emission peaks ($\sigma^+$ and $\sigma^-$), which means that the VHE can still be seen at room temperature. The separation of PL peaks is $0.30 \pm 0.02\,\mu m$, which is slightly smaller than the case in low temperature.

To check the robustness of the VHE in our experiment, we have examined the PL emission patterns under different conditions. First, we changed the excitation spot to different locations on the samples (shown in Fig. 4d-f). While the emission pattern is different (Fig. 4a vs Fig. 4d), the $\sigma^+$ intensity peak always shows in the right side of the transport direction, following the same rule in our observed VHE. Next, we have used different excitation polarization (Fig. S6 and Fig. S7). The excitation with horizontal linear polarization is very similar with the case of the vertical one as shown in Fig. 3. Away from the excitation spot, the VHE can still be observed. Furthermore, we have demonstrated the VHE with different excitation laser wavelengths (See Fig. S8) which shows that the observed VHE is robust against excitation laser energy. Finally, in order to validate the potential trapping effect caused by the strain and exclude the possible affecting of the photonic crystal substrate, we fabricate new samples on substrate with only strain at the edge of the Silicon-On-Insulator (SOI) substrate. The exciton funneling is clearly observed on the strained area too and interlayer exciton VHE is demonstrated (supplementary Fig. S10).

In summary, we report the room temperature interlayer exciton Hall effect in MoS2/WSe2 heterostructures. The robustness of exciton transport at room temperature might come from the much longer exciton and valley lifetimes as compared to that of intralayer excitons. Future work may include using electric field to control the transport of interlayer excitons since they possess an out of plane dipole with electron and hole separated in different layers[25-27]. Other fundamental questions include how the exciton repulsive interaction induced valley-polarized exciton gas and valley Hall effect affect each other; and how the VHE plays a role in exciton superfluidics in TMD heterostructures[35]. We note that valley separation has been achieved with with the assistance of well-designed metasurface[36] and subwavelength asymmetric groove array[37]. Our demonstration of the control of room temperature valley information might open new possibilities of opto-electronics and valleytronics application for interlayer excitons.

## Methods

**Sample fabrication.**

The monolayers of WSe2 and MoS2 were mechanically exfoliated using viscoelastic polymer (polydimethylsiloxane), and then deposited consecutively onto a boron nitride flake on the silicon photonic crystal slab in a dry procedure, with their stacking angle was near zero degrees. The sample was then transferred to a chamber and annealed with Ar/H2 (95%/5%) gas blowing at 473K for 3 hours, to improve the contact. The monolayers were optically selected with a microscope and confirmed by PL and Raman signal.

**Polarization resolved PL mapping.**

The polarization and spatial resolved PL measurement was carried out by a home-made confocal microscope, as shown in Fig. 2a. Our sample is positioned by a three-dimensional piezoelectric stepper, which is mounted on a closed-cycle optical cryostat. The 726 nm diode laser is focused onto the sample with beam diameter around 1 $\mu$m, using a 50x infrared objective lens. The PL emission from the interlayer excitons was selected using a 1050 nm long pass filter. By rotating the polarizers and wave plates, we can reach any specific polarization state. To get Fig. S1and Fig.S3, we first tune Galvo 2 to let the detection spot meets the excitation spot .Then we use Galvo 1 to perform the PL mapping by moving that spot across the sample. For all other mappings, we fix Galvo 1 and therefore the excitation spot on the sample, while at the same time scan Galvo 2 placed at the detection arm. The detection fiber core will be imaged onto different locations on the sample by moving Galvo 2, therefore a PL emission profile can be obtained. With the beam splitter and a multi-channel SSPD, the $\sigma^+$ and $\sigma^-$ components of the PL signal from each spot were measured simultaneously. The systematic efficiency difference of these two channels is cancelled by combining the data before and after tuning the half wave plate at 45 degrees on the detection arm, and therefore switches the $\sigma^+$ and $\sigma^-$ for the two detection channels.

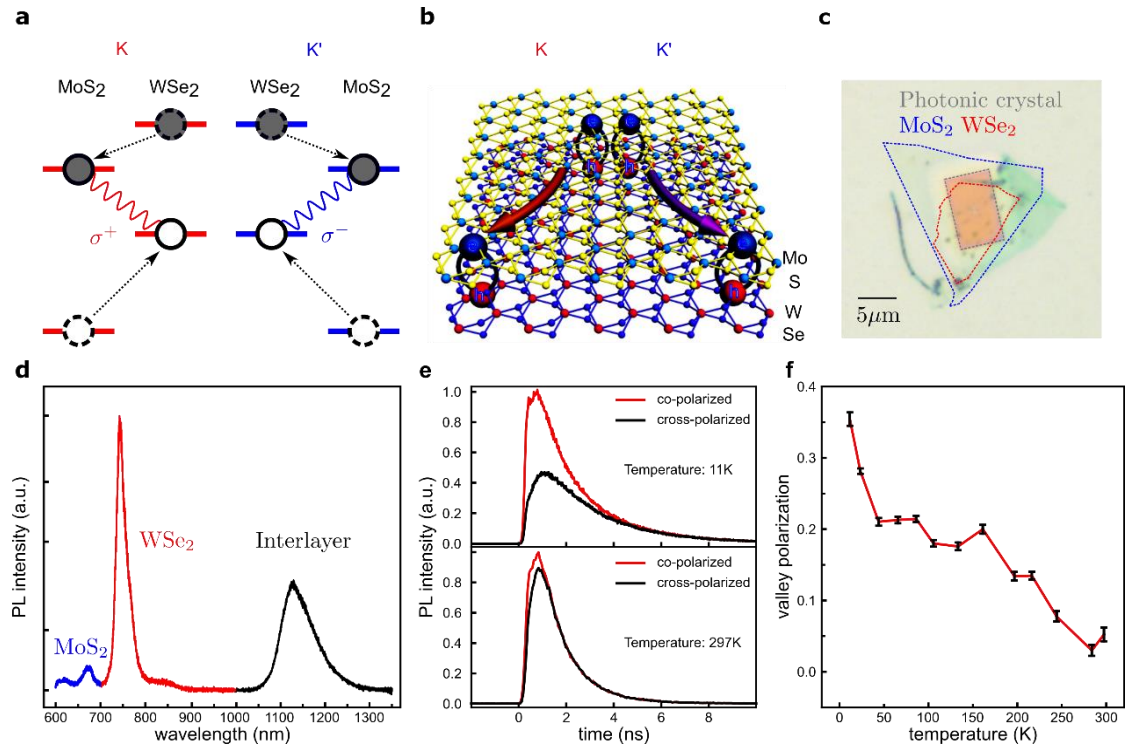

**Figure 1. Concept of interlayer exciton valley hall effect. a**, Interlayer exciton in different valleys. Electrons (grey dots) will hop to MoS2 with a lower energy in the conduction band. Holes (empty dots) will hop to the valence band of WSe2. The combination of electrons and holes emits photons with optical helicity. $\sigma^+$ ($\sigma^-$) circularly polarized light couples to the K (K') valley. **b**, Side view of heterostructures with MoS2 and WSe2 monolayers. Interlayer excitons formed by electrons (e) and holes (h) will traverse to different directions in different valleys (K and K'). **c,** Microscope image for our sample. Red dashed line represents monolayer WSe2. Blue dashed line represents monolayer MoS2. WSe2 and MoS2 are stacked on a silicon-on-insulator (SOI) substrate by dry transfer. In the middle of the substrate, a photonic crystal suspended slab is fabricated in rectangle shape. **d**, PL from intralayer and interlayer excitons. Wavelength of interlayer exciton emission is well separated from that in monolayer MoS2 and WSe2, making sure we can detect interlayer exciton emission exclusively by using a long pass filter. **e**, Time-resolved PL measurement for co-polarized and cross-polarized situation in low temperature (up) and room temperature (down). **f**, Valley polarization as a function of temperature. The valley polarization starts from around 35% and reduce to about 5% as the temperature goes higher form cryogenic temperature to room temperature.

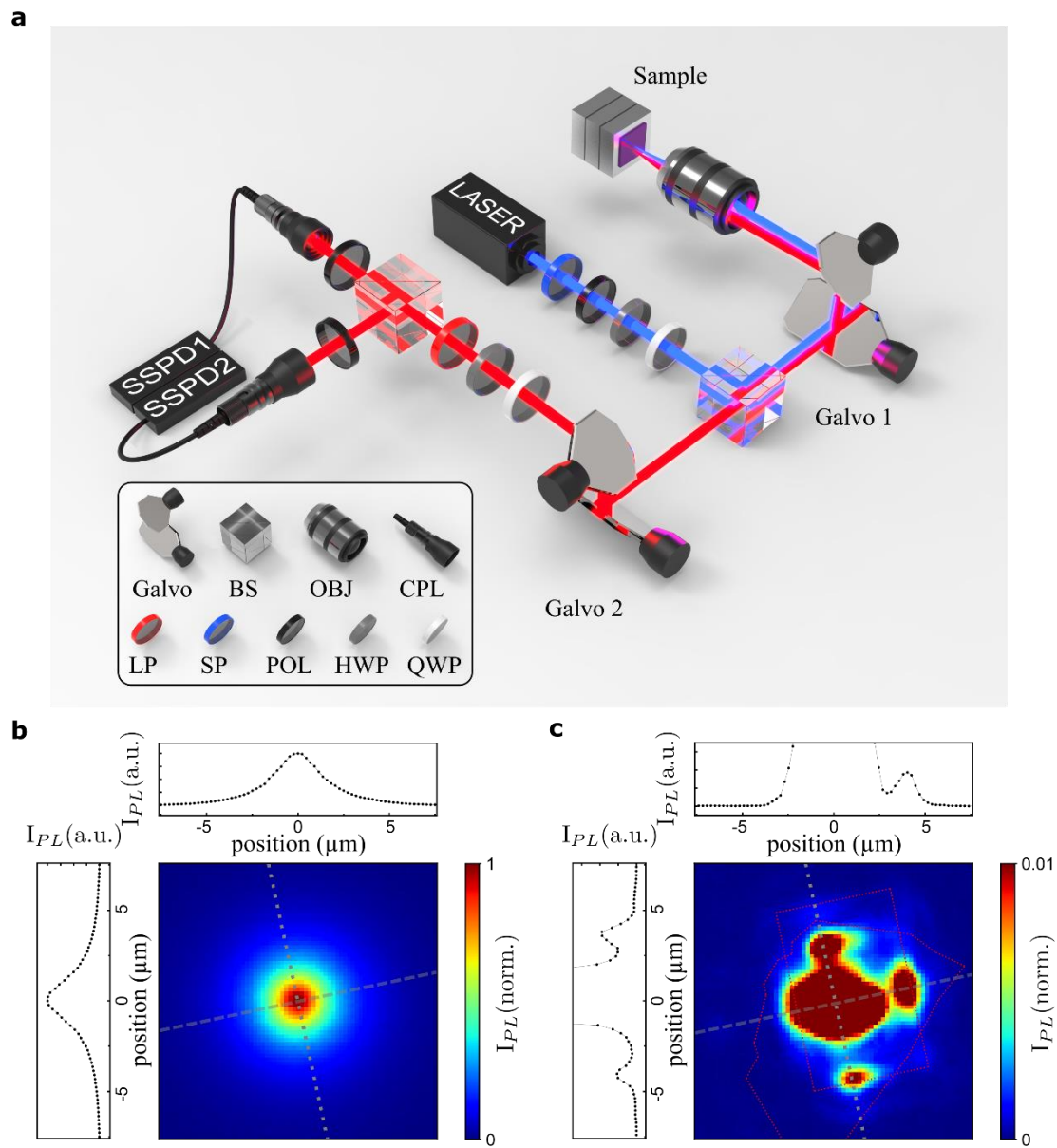

**Figure 2. Experimental setup and detection of exciton transport. a**, Experimental setup. The sample is in a closed-cycle cryostat with variable temperature. A laser with controllable polarization is sent to the sample through a scanning galvo mirror set. PL is sent to the superconducting single photon detector (SSPD). Galvo in the PL collection arm is used to implement PL mappings (See detail in the Method). BS, beam splitter; OBJ, objective lens; CPL, coupler; LP, long pass filter; SP, short pass filter; POL, polarizer; HWP, half wave plate; QWP, quarter wave plate. **b**, The emission pattern for silicon PL. By scanning the Galvo mirror set, we are able to detect the emission spatial profiles. As expected, the emission spot shows a Gaussian profile. Intensity as a function of position is shown on top for the horizontal linecut. The one for the vertical linecut is shown on the left. **c**, Emission pattern for interlayer exciton PL. In stark contrast, here the emission shows a non-Gaussian profile. To see the profile more clearly, here we set the maximum intensity as one percent of the observed intensity

peak. From the intensity of linecuts, we can see photon sidebands in both horizontal and vertical directions. These sidebands come from the exciton transport in a sizable distance. Edges of heterostructures sample and suspended silicon slab are drawn in the red dashed line. Local changes on these edges can serve as excitons traps. They will enhance the emission from at these edges and introduce asymmetry for the exciton transport. Glavo, Galvo 2D scanning mirrors system; BS, beam splitter; OBJ, objective; CPL, coupler; LP, longpass filter; SP, shortpass filter; POL, polarizer; HWP, half-wave plate; QWP, quarter-wave plate.

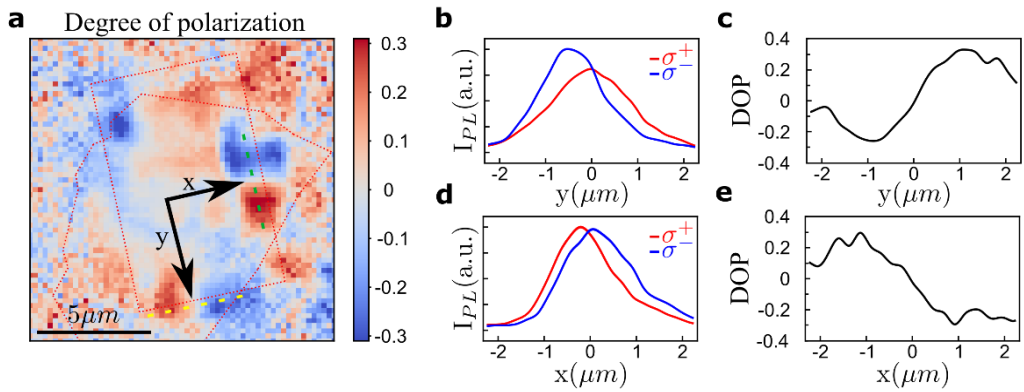

**Figure 3. Experimental demonstration of exciton valley Hall effect in cryo-temperature. a**, Degree of polarization. The origin of the x and y axes is at the excitation spot. The edges of heterostructures and suspended rectangle substrate are shown with the red dashed line (the same for all other figures). **b,** PL intensity with different polarization following the green linecut in **c**. A separation of the intensity peak can be seen in the y direction, perpendicular to the exciton transport in the x direction. **c,** Degree of polarization (DOP) as calculated from **b**. **d,** PL intensity with different polarization following the yellow linecut in **a**. Separation of emission intensity peak and therefore different exciton movement in the transverse direction can be seen as well. **e,** DOP as calculated from **d**.

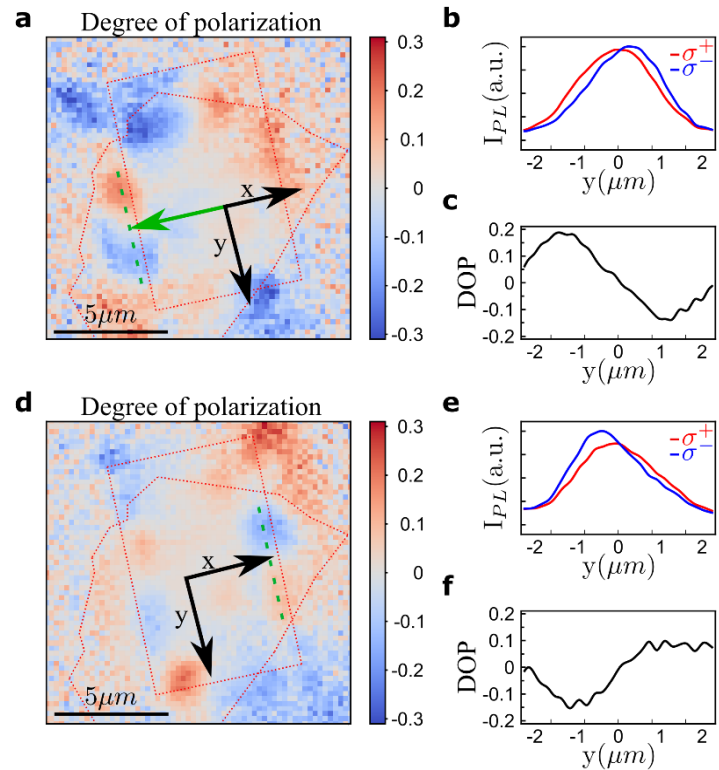

**Figure 4. Room-temperature valley hall effect. a,** Degree of polarization. The green arrow represents the exciton transport direction from the excitation spot. **b, c,** Intensity and DOP following the dashed green linecut in **a**. **d,** Degree of polarization with another excitation spot. **e, f,** Intensity and DOP following the dashed green linecut in **d**. The intensity peak separation is reversed in **a** and **d**, due to the reversal of the dominant exciton transport direction.

# Supplementary Information for Room temperature interlayer exciton valley Hall effect

**This file includes:**
   **A. MoS2/WSe2 interlayer excitons**

   **B. The influence of the silicon photonic crystal slab**
   **C. Setup characterizations**
   **D. Theoretical simulation for the exciton pattern**
   **E. Excitation polarization dependence**
   **F. Excitation wavelength dependence**
   **G. VHE experiment on sample II with photonic crystal substrate**
   **H. VHE experiment on sample III with etched structure on substrate**

## A. MoS2/WSe2 interlayer excitons

For the MoS2/WSe2 interlayer excitons, we observe the photoluminescence peak at ~ 1.05 eV, which is attributed to K-K transitions, consistent with previous STM measurement[38]. This is different from the previous observation for the $K - \Gamma$ transition around 1.6 eV[31, 32]. Figure S1 shows a typical heterostructure sample and PL mapping. The PL mapping shows that the emission peak at 1150 nm is only in the heterostructure region. This excludes the possibility of such emission coming from defect emission. The above observation has been repeated for multiple samples.

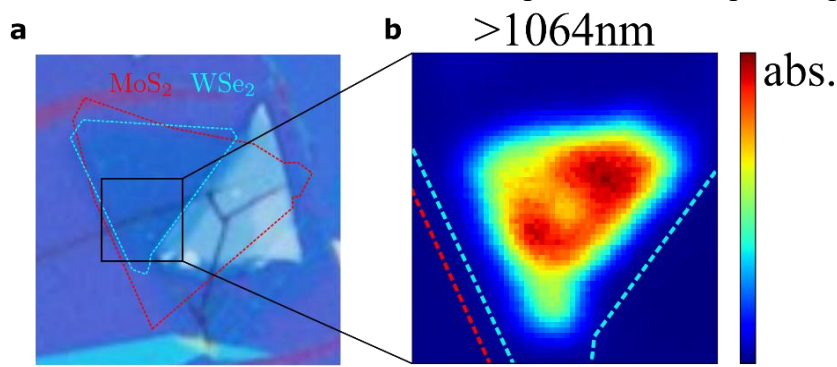

**Figure S1. Optical image of one heterostructure sample and spatial distribution of PL with wavelength above 1064nm. a,** the optical image of a MoS2-WSe2 heterostructure sample. **b,** PL mapping for the black square region shown in **a**. This shows the PL emission above 1064 nm is only form a corner of the heterostructure region.

## B. The influence of the silicon photonic crystal slab

The photonic crystal suspended slab helps for observing the interlayer exciton VHE. For MoSe$_2$-WSe$_2$ heterostructure on SiO$_2$ substrate, interlayer excitons of both valleys have a Gaussian shaped drift-diffusion. As mentioned in the main text, our photonic crystal slab underneath can create local traps at the edge to modify the exciton transport path. Excitons will have a non-centrosymmetric distribution that introduces the longitudinal movement of the interlayer excitons.

Another important help is that PL is significantly enhanced at the part on the slab. For photonic crystal substrate, the lattice constant of the structure is $a$ = 296.2 nm and a radius of the air holes $r$ = 83.4 nm. This combination of geometries leads to a relative thickness of $d/a$ = 0.74 and relative hole size of $r/a$ = 0.282. We adopt 3D FDTD solutions to calculate the bandgap of this photonic crystal structure. In simulation, a group of dipoles with various orientations is used to excite the possible modes in this structure. The calculated bandgap structure of TE modes is shown in Figure S2. As can be seen, it has a photonic bandgap ~240nm and therefore serves as a mirror for PL range around 1150nm. Indeed, an enhancement of PL emission is observed as shown in Figure. S3.

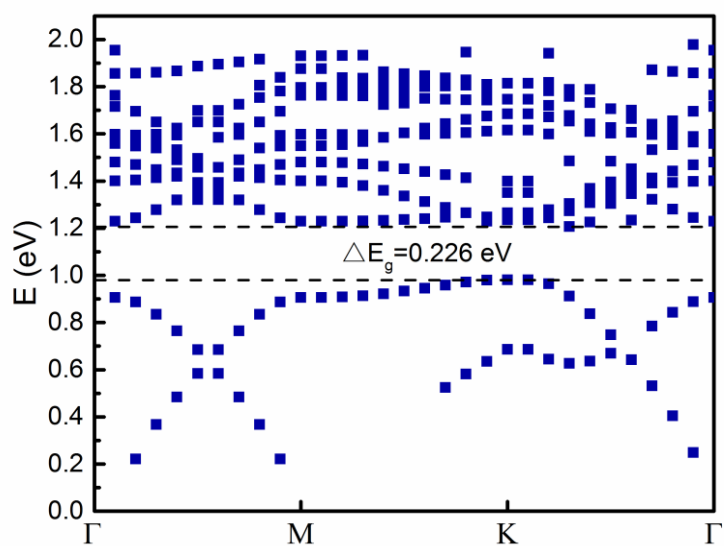

**Figure S2. Band structure of the photonic crystal structure for the TE modes.** The blue dots represent TE bands and the dashed black lines represent the photonic bandgap of this structure is around 0.266 eV.

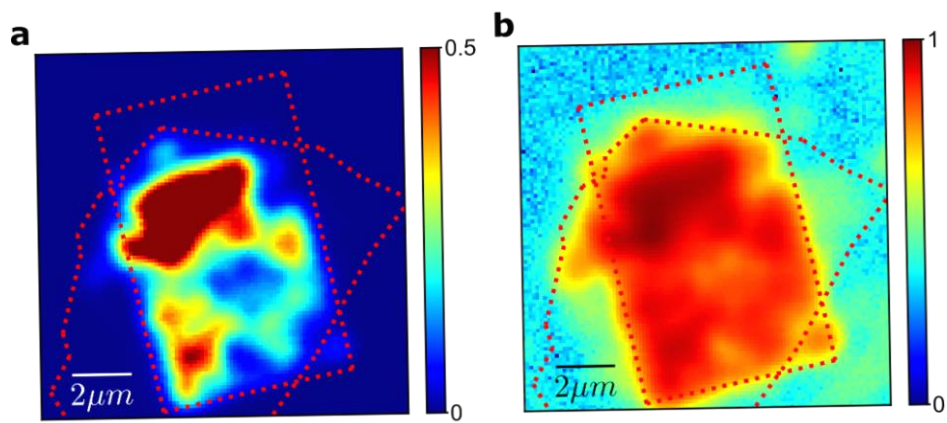

**Figure S3. Color plot of interlayer exciton PL intensity mapping. a, b,** The linear

and log plot of PL from MoS2/WSe2 heterostructure, respectively. Here the excitation spot is scanned across the sample. The dotted red lines show the outline of the photonic crystal and heterostructure.

## C. Setup characterization

To measure the polarization-dependent PL mapping, our two collection arms need to be precisely aligned. For Photoluminescence (PL) signals without the Hall effect, the PL mapping should be the same for $\sigma^+$ and $\sigma^-$ detection. Instead of using the laser reflection signal, whose wavelength is different with the interlayer exciton emission, we choose the Silicon PL as a reference. It has a wavelength centered around 1100 nm, which is close to our interlayer exciton emission. Even though its intensity is ~100 smaller than the interlayer excitons, it is enough to calibrate our setup.

We moved out of the heterostructure area and focused on a substrate without heterostructure samples. Subsequently, by focusing the excitation arm on the sample and then scanning the detection map of the silicon PL, the data shows Gaussian shape for both $\sigma^+$ and $\sigma^-$ detection (Fig. S4a and S4b). Ideally the Degree of polarization (DOP) calculated using these two sets of data should be exactly zero. For our experimental setup, the DOP for silicon PL is within ±3% . We note that this is much smaller that the measured interlayer exciton DOP, which is around 30 %. In conclusion, our setup is aligned precisely enough to measure the interlayer exciton VHE.

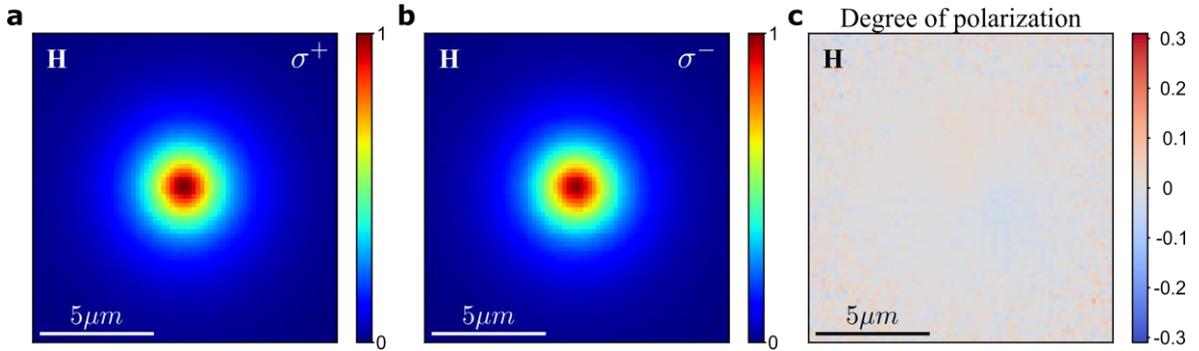

**Figure S4. Spatial distribution of Silicon PL signal in SiO2-Si substrate. a, b,** Spatial profile of emitting silicon PL signal under H linear excitation, and $\sigma^+$ and $\sigma^-$ detection, respectively. The signal is taken at the blank part of another heavily doped SiO2-Si substrate. **c,** Degree of polarization calculated for silicon signal using data form **a** and **b**. For DOP, we use the same color bar as the main Fig. 3a and Fig.4a, 4d. A nearly white image shows that the experimental setup is well aligned with negligible DOP for silicon PL.

## D. Theoretical simulation for the exciton pattern

As discussed in the main text, the excitons in different valleys carry opposite Berry curvatures. One consequence is that when encountering a potential gradient, the excitons will gain a valley dependent transverse velocity. A way to describe such an effect is to introduce a term in the exciton velocity[39]:

$$v_\Omega = \overrightarrow{grad}V(r) \times \vec{\Omega}$$

where $\Omega$ is the Berry curvature and V is the potential.

In order to study the dynamics of the excitons, we choose to solve iteratively the equation of motion associated to the exciton ballistic motion and then calculate the average trajectory of many excitons, which are all assumed to originate at the pump spot center with uniformly distributed initial propagation direction and Fermi-Dirac distribution of initial energies (corresponding to different initial speeds).

Since obviously the distance between the center of the pump and the edge depends on the direction, the intensity of the exciton will be larger for four particular directions, accounting for a square confinement potential. Therefore for a centered pump we expect to obtain 5 high intensity spots, the central pump and 4 recombination spots on the edge. By reaching the edge of the system, the excitons encounter a potential barrier formed by the interface between the sample and the substrate. This kind of potential can for example be explained by a local deformation of the lattice. It has the effect of slowing down the excitons and creates a transverse valley dependent velocity for the excitons as discussed before. The localized pump laser can induce a temperature gradient from the center of the sample to the edge[2]. However this gradient seems to be very small at room temperature and therefore we choose to not include it in the simulations.

Introducing the edge potential in the equation of motion along with the Berry induced velocity, we obtain the DOP in the vicinity of the edge of the sample. The DOP strongly depends on the value of the Berry curvature and on the intensity and shape of the potential. We here take the potential as half a Gaussian near the edge with a maximum amplitude of 200 meV and a half width of 0.5 microns. Similar as TMD monolayers, here we take the Berry curvature in TMD monolayer as 15 Å². Consequently, since we take the classical value for the Berry curvature, the DOP values are smaller than in the experiment. The difference between the intensity distributions in the model and the experiment can be explained by an extra energy relaxation term. With those parameters, we obtain the DOP and intensity distributions shown in Fig. S5, for central pump on a square sample of 10 by 10 microns at 270K:

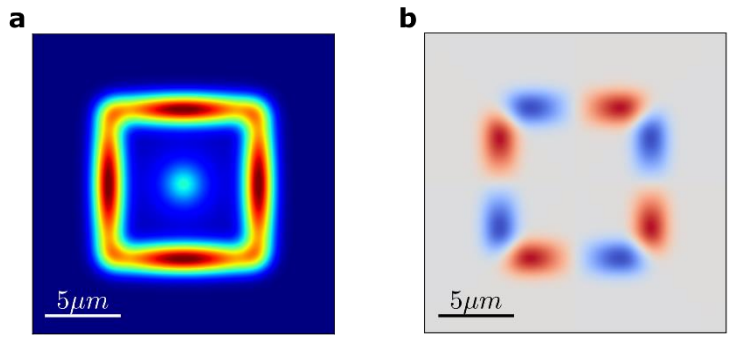

**Figure S5. Degree of polarization and intensity distribution in the simulation. a**, Simulated total emission from a square shaped edge potential area. **b**, simulated DOP distribution. The $\sigma^+$ and $\sigma^-$ emission will split along the edge because of VHE, and build up DOP difference.

### E.   Excitation polarization dependence

Except for the vertical linear polarization excitation, our sample is also examined using Horizontal, $\sigma^+$ and $\sigma^-$ polarized excitation for both cryogenic and room temperature. What our result shows is that, no matter what excitation polarization is used, all the excitons transported to the edges follow the same rule: K valley excitons turn to right and K' valley excitons turn to left. For cryo-temperature horizontal linearly polarized excitation (Fig. S6), the result is almost the same as vertical linearly polarized excitation (Fig. 3). In room temperature (Fig. S7), even the intervalley scattering is significantly enhanced and valley polarization is tiny, the excitons are still separating along the direction perpendicular to their transporting direction (Fig. S7 b and e), showing strong robustness of VHE against excitation polarization and temperature.

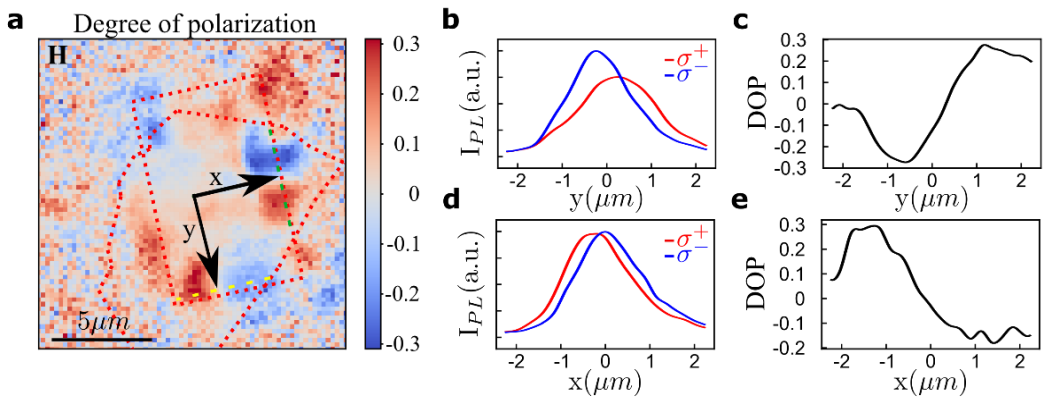

**Figure S6. The interlayer exciton VHE at cryogenic temperature. a,** Color plot of the interlayer exciton DOF using horizontal linearly polarized excitation. Two directions (right as green and down as yellow) show a clear splitting of the excitons with different polarizations. **b**, **c (d**, **e)**, $\sigma^+$, $\sigma^-$ PL and DOP at green (yellow) linecut, which show interlayer exciton VHE.

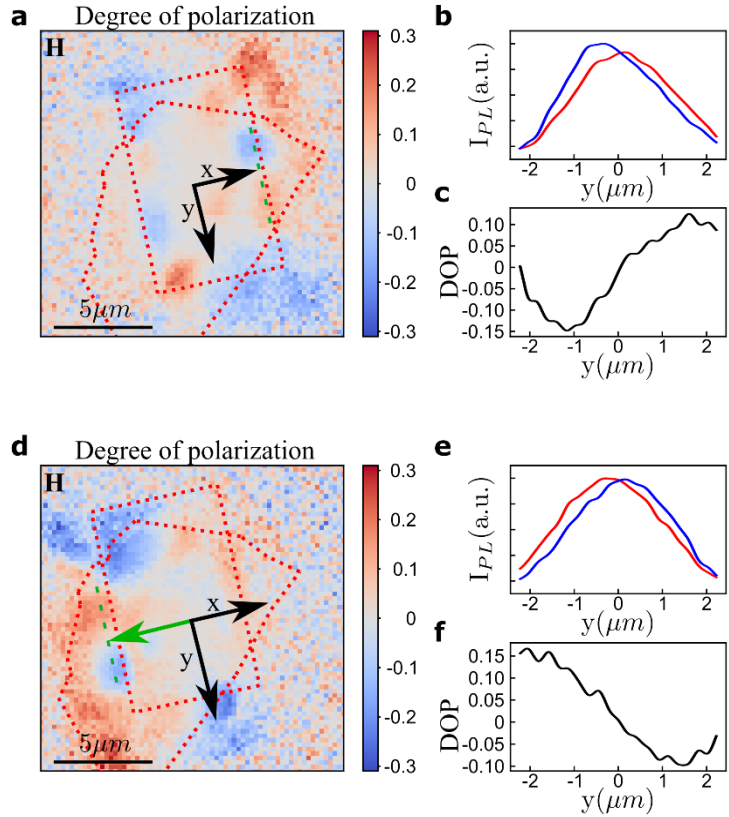

**Figure S7. The interlayer exciton VHE at room temperature. a,** Color plot of the exciton hall effect using horizontal linearly polarized excitation. The excitons that move right show a clear splitting in their trajectories with different polarization. **b, c,** $\sigma^+$, $\sigma^-$ PL and DOP linecut at the green line. **d-f,** same with **a-c** but with another excitation spot which have interlayer excitons move left.

## F. Excitation wavelength dependence

Besides a 726nm CW laser, we also tested the performance of the interlayer exciton VHE using 532 nm and 690 nm diode laser excitation. Excitons that move in the x direction (right) are convincing evidence to show the interlayer exciton VHE robustness with different excitation wavelengths.

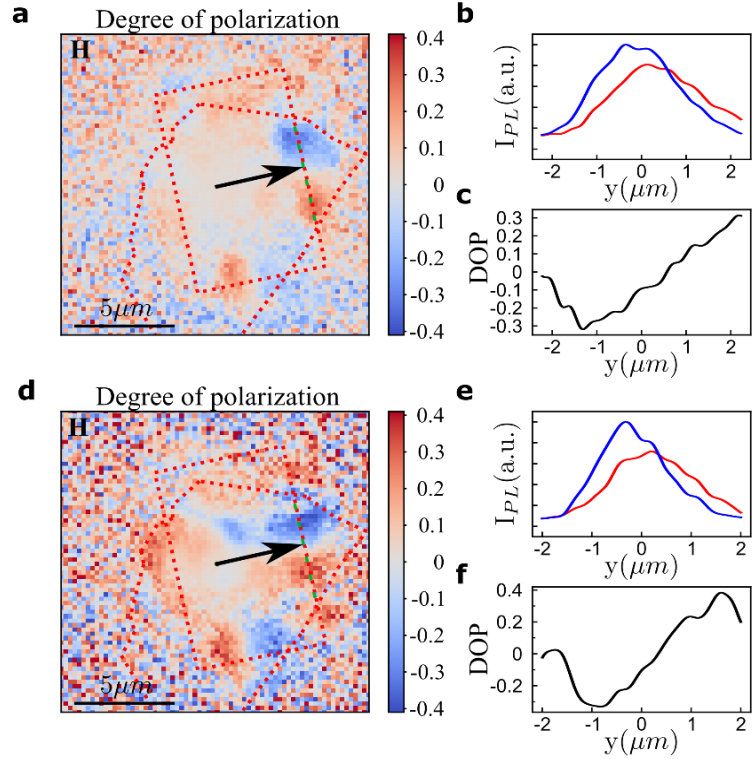

**Figure S8. Interlayer exciton VHE excited with different wavelength. a,** DOP profile when pumped with 532 nm laser at horizontal linearly polarized excitation. **b, c,** $\sigma^+$, $\sigma^-$ PL and DOP linecut at the edge (green line in **a**). **d-f,** same with **a-c** but pumped with 690 nm laser at horizontal linearly polarized excitation.

### G. VHE experiment on sample II with photonic crystal substrate

We fabricated another sample (sample II) to check the exciton VHE too. Here we excite at the bottom of the sample and emission on the upper side of the sample can be seen, which shows a separation of two circulation polarization peaks. In Fig. S9 (a-c), excitation light has a right circular polarization $\sigma^+$. In Fig. S9 (d-f), excitation light has a left circular polarization $\sigma^-$.

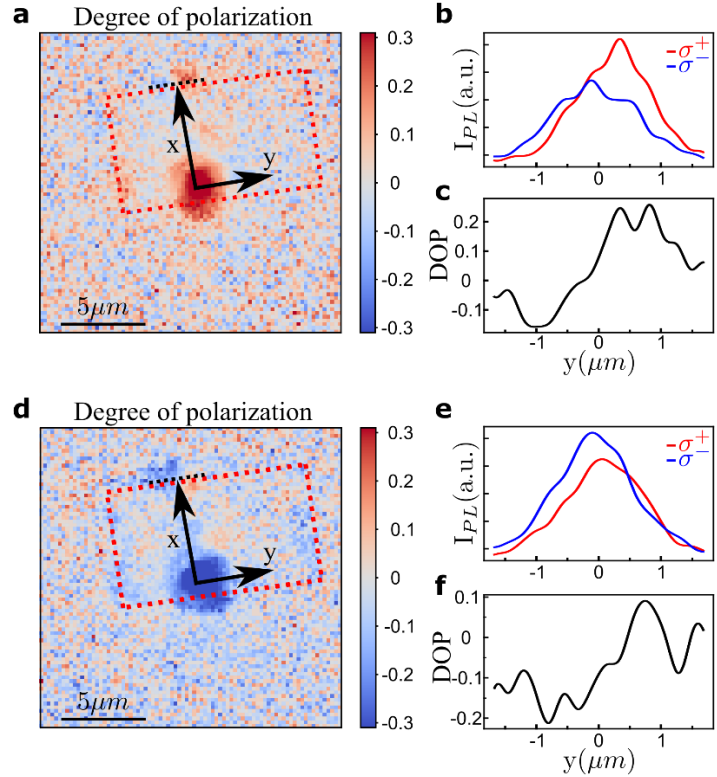

**Figure S9. Interlayer exciton VHE shown in another sample with photonic crystal substrate underneath**. The structure of the photonic crystal is similar with the one in main text. **a (d),** the DOP profile of interlayer exciton using $\sigma^+$ ($\sigma^-$) excitation. We excite the bottom region of the photonic crystal, and at the top region we see a separation of red and blue, with blue to left side and red to right side. **b (e),** PL intensity at the dotted black line in top region. **c (f),** DOP calculated from b (e).

## H. VHE experiment on sample III with etched structure on substrate

In order to validate the potential trapping effect caused by the strain and exclude the possible affecting of the photonic crystal substrate, we fabricated a new batch of samples without photonic crystal structure and only holes at the edge of the SOI substrate. Sample images after fabricating MoS2/WSe2 heterostructure are shown in Fig. S10a. Then we focused light in the center of the sample and scanned the detection area. As can be seen from Fig. S10b, even though only the sample center is excited, emission can come from the sample edge in the strained area. This clearly shows the exciton funneling effect[33], which agrees with experiment demonstration of strain effect as shown in previous works[34]. Under linear polarization excitation and with polarization-resolved emission pattern, DOP can be calculated in Fig. S10c. Excitons in one valley turn to the transverse direction perpendicular to the exciton propagation direction, causing DOP<0 at the left side and DOP>0 at the right side following the propagation directions. PL intensity and DOP along two cutting lines in Fig. S10c are shown in Fig. S10d-g.

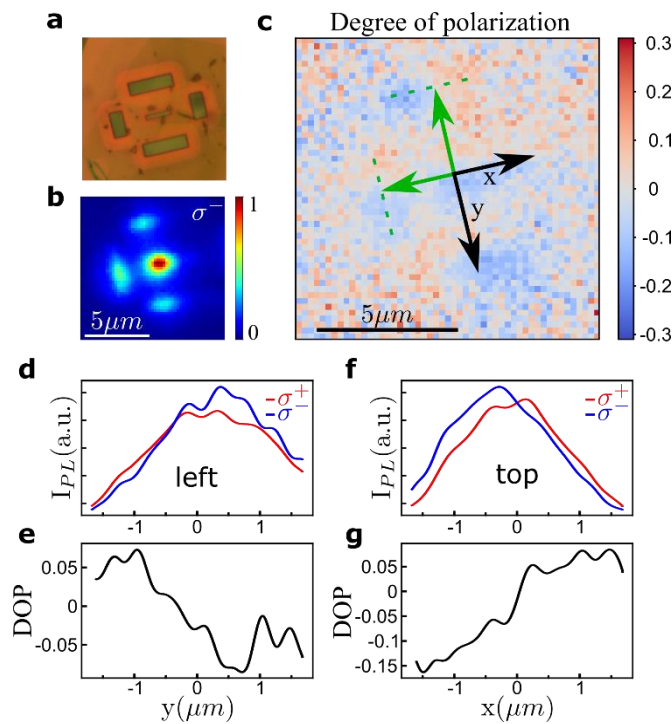

**Figure S10. Interlayer exciton VHE on another sample. a,** optical image of samples after fabrication of TMD on the substrate with etched rectangle-shape holes. **b,** The diffusion of excitons. **c,** Color plot of the exciton Hall effect. The center point represents the excitation points and green arrow represents the exciton diffusion direction. **d, e (f, g),** $\sigma^+$ and $\sigma^-$ PL intensity separation along y (x) direction as the excitons are moving in -x (-y) direction, and the DOP calculated form d (f), respectively.